\documentclass[RNAAS]{aastex631}

\received{5 October 2021}
\accepted{12 October 2021}
\submitjournal{RNAAS}

\shorttitle{A possible very recent ejection in the Trapezium}
\shortauthors{Ma{\'\i}z~Apell\'aniz et al.}

\begin{document}

\title{$\theta^1$~Ori C as a medieval bully: a possible very recent ejection in the Trapezium}

\correspondingauthor{J. Ma{\'\i}z Apell\'aniz}
\email{jmaiz@cab.inta-csic.es}

\author[0000-0003-0825-3443]{J. Ma{\'\i}z~Apell\'aniz}
\affiliation{Centro de Astrobiolog\'{\i}a, CSIC-INTA, Spain}

\author[0000-0001-9933-1229]{M. Pantaleoni~Gonz\'alez}
\affiliation{Centro de Astrobiolog\'{\i}a, CSIC-INTA, Spain}
\affiliation{Universidad Complutense de Madrid, Spain}

\author[0000-0003-1086-1579]{R. H. Barb\'a}
\affiliation{Universidad de La Serena, Chile}




\begin{abstract}
We use {\it Gaia} EDR3 astrometry to propose that a dynamical interaction between the multiple system $\theta^1$~Ori~C and
$\theta^1$~Ori~F ejected the latter as a walkaway star $\sim$1100~years ago (without deceleration) or somewhat 
later (with a more likely deceleration included). It is unclear whether the final 3-D velocity of $\theta^1$~Ori~F will be large 
enough to escape the Orion nebula cluster.
\end{abstract}



\section{Introduction} \label{sec:intro}

$\,\!$\indent Stars can be ejected by dynamical interactions among 3+ bodies in stellar clusters \citep{Poveetal67}. 
Such objects are called runaway stars when their space velocity is $>30$~km/s \citep{Hoogetal01}
and are being found in increasing numbers with {\it Gaia} astrometry \citep{Maizetal18b}.
Walkaway stars, ejected with velocities $<30$~km/s, are even more common than runaways \citep{Renzetal19b}.

The prototype dynamical ejection happened near the Orion nebula 2.5~Ma ago, resulting in the 
high-speed expulsion of AE~Aur and $\mu$~Col \citep{BlaaMorg54} and of a third multiple system, $\iota$ Ori
\citep{MaizBarb20}, moving more slowly \citep{Hoogetal01}. $\theta^1$~Ori~Ca, the main component in 
$\theta^1$~Ori~C, is a magnetic Of?p object \citep{Maizetal19b}, the most massive star in the Orion nebula
cluster, and part of a young binary system with an 11~years period and a possible third member \citep{Krauetal09b,Lehmetal10}. 
$\theta^1$~Ori~C may have caused the ejection of the Becklin-Neugebauer object from the Trapezium 4~ka ago 
\citep{Tan04} but it is likely too young to have been involved in the previous event. Recent papers have
proposed other walkaway and runaways from the Orion nebula cluster \citep{Schoetal20,Farietal20,Platetal20,Maizetal21e}.

\section{Methods} \label{sec:methods}

$\,\!$\indent In the Villafranca project \citep{Maizetal20b} we have analyzed the Orion nebula cluster and 
determined a distance of 390$\pm$2~pc and proper motions of $1.577\pm0.022$~mas/a in $\alpha$ and $0.464\pm0.022$~mas/a in $\delta$ 
\citep{Maizetal21e}. Here we analyze the 
{\it Gaia}~EDR3 astrometry of two Trapezium objects, $\theta^1$~Ori~C and $\theta^1$~Ori~F. We 
apply the zero point of \citet{Maiz21} to the parallaxes and correct the proper motions according 
to \citet{CanGBran21}. The astrometric external uncertainties ($\sigma_{\rm ext}$) are calculated with the $\sigma_{\rm s}$  
of \citet{Maizetal21c} and \citet{Lindetal21a} and the $k$ of \citet{Maiz21}.

\begin{figure}
\centerline{\includegraphics[width=1.02\linewidth]{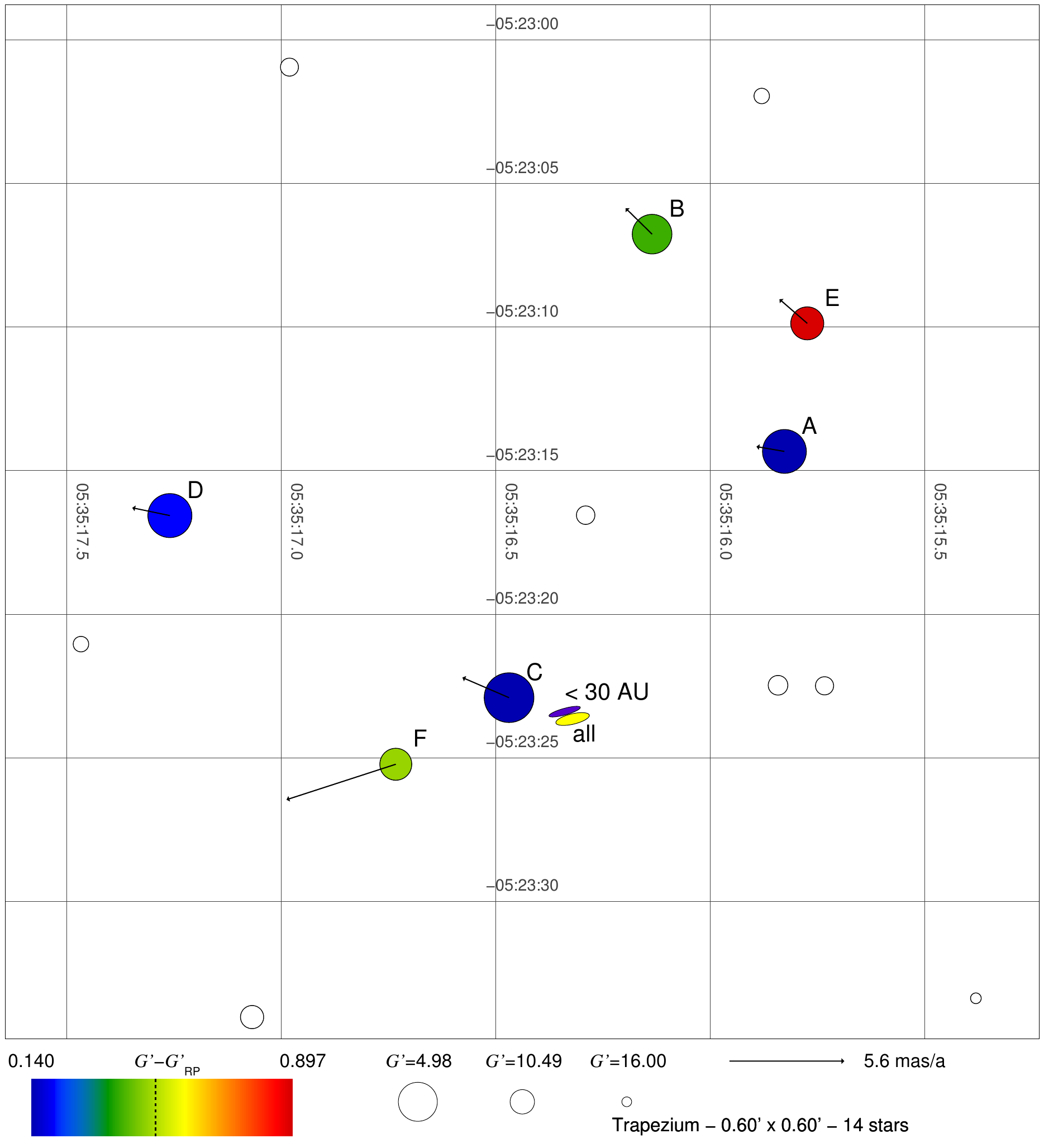}}
\caption{{\it Gaia}~EDR3 chart of the central 36\arcsec$\times$36\arcsec\ of the Trapezium. Circle size encodes 
         magnitude and color encodes $G^\prime-G^\prime_{\rm RP}$, with empty circles for objects with poor color 
         information. Arrows represent proper motions. 
         The yellow and violet ellipses indicate the position and one-sigma uncertainty ellipses in the plane of the 
         sky of the minimum distance from the Montecarlo simulations, with the first one used for all results and the second one 
         for those that have a minimum distance of less than 30 AU.}
\label{Trapezium}
\end{figure}

\vfill

\eject

\section{Results} \label{sec:results}

\begin{itemize}
 \item The five brightest stars in $\theta^1$~Ori (A to E) have proper motions similar to that of the cluster. F is an outlier.
 \item The parallaxes of $\theta^1$~Ori~C and $\theta^1$~Ori~F are within 0.2 $\sigma_{\rm ext}$ of the cluster parallax.
       Furthermore, their differences in $G^\prime$ and $G^\prime_{\rm BP}-G^\prime_{\rm RP}$ are similar to those expected from their differences 
       in spectral types (see below). Therefore, they are at similar distances and do not experience very different extinctions, 
       contradicting the hypothesis by \cite{Olivetal13} that $\theta^1$~Ori~F is a foreground object.
 \item The movement in the plane of the sky as traced by the {\it Gaia}~EDR3 proper motions of the two stars point towards a common
       region of the sky (Fig.~\ref{Trapezium}).
 \item We have carried out Montecarlo simulations of the trajectories and determined that the minimum 
       distance (in the plane of the sky) took place 1.10$\pm$0.07~ka before the {\it Gaia}~EDR3 epoch (2016.0) with a value of 
       340$\pm$110~AU at a position of $\alpha=$05:35:16.321, $\delta=-$05:23:23.65 (J2000) with uncertainties of 0\farcs394 and 
       0\farcs119 along axes at PAs of $-77^{\rm o}$ and $13^{\rm o}$.
 \item The tail of the Montecarlo simulations reaches a minimum distance of zero. Selecting the subset where the minimum distance 
       is less than 30~AU we obtain a very similar flight time to 2016.0 of 1.09$\pm$0.07~ka and a slightly different position 
       (Fig.~\ref{Trapezium}) of $\alpha$~=~05:35:16.339, $\delta$~=~$-$05:23:23.39 (J2000) with uncertainties of 0\farcs372 and 
       0\farcs076 along axes at PAs of $-75^{\rm o}$ and $15^{\rm o}$.
 \item Looking in the WDS \citep{Masoetal01}, separations decrease as we go back in time (the first observations are from the 
       nineteenth century) with some scatter and a zero value compatible with an epoch 600-1200~a ago.
 \item We have no GOSSS \citep{Maizetal11} or LiLiMaRlin \citep{Maizetal19a} spectra of $\theta^1$~Ori~F. \citet{Cost19} 
       classifies it as a chemically peculiar B7.5~p~Si star, suggests a spectroscopic mass of 3.7~M$_\odot$, and determines that
       its radial velocity in the frame of reference of $\theta^1$~Ori~C is $6.2\pm 4.2$~km/s towards us.
 \item The relative velocity between $\theta^1$~Ori~C and $\theta^1$~Ori~F in the plane of the sky is 7.8$\pm$0.5~km/s. Adding
       the radial velocity leads to a 3-D velocity of 10.0$\pm$2.6~km/s. Therefore, $\theta^1$~Ori~F is at 
       most a walkaway star, not a runaway. However, caution is needed because discrepancies of several km/s in the measurements
       of radial velocities of OB stars are common \citep{Trigetal21}.
 \item The two systems are currently separated by 1780~AU in the plane of the sky. Assuming a total mass of $\sim$50~M$_\odot$
       \citep{Krauetal09b,Lehmetal10} and that the trajectory is mostly in the plane of the sky leads to a escape velocity of 
       $\sim$7~km/s. For a 3-D trajectory, the value would go down to $\sim$5~km/s. As that value is just $\sim1/2$ of the measured
       velocity, the assumption above that the trajectory is linear is wrong. The system must have experienced a 
       significant deceleration since maximum approach (hence reducing the flight time) and should slow down even more in the
       future. Furthermore, the chances of a close minimum distance are increased.
\end{itemize}

\section{Conclusions} \label{sec:conclusions}

$\,\!$\indent These results indicate that $\theta^1$~Ori~F was ejected from the $\theta^1$~Ori~C 
(currently double or triple) system during the middle ages (1.1~ka ago or less including deceleration). This is 
a recent example of the complex three-or-more-body encounters that take place in very young stellar clusters with massive stars
such as the one in the Orion nebula. Nevertheless, the small difference between the measured velocity and the
escape velocity from $\theta^1$~Ori~C does not allow us to clearly establish whether $\theta^1$~Ori~F will leave the cluster after
escaping from $\theta^1$~Ori~C or not \citep{Costetal08}. See \citet{Maizetal17a} for another example of a recent dynamical encounter 
between (very) massive stars and with an also uncertain outcome.

\begin{acknowledgments}
J.M.A. and M.P.G. acknowledge support from the Spanish Government MCIn through grant 
PGC2018-095\,049-B-C22. R.H.B. acknowledges ANID FONDECYT Regular Project 1\,211\,903.
\end{acknowledgments}

\vfill

\eject

\end{document}